\documentclass[twocolumn]{aastex63}

\submitjournal{ApJ}

\accepted{\today}

\shorttitle{Disk Height}
\shortauthors{Rich et al.}

\graphicspath{{./}{figures/}}

\begin{document}

\title{Investigating the Relative Gas and Small Dust Grain Surface Heights in Protoplanetary Disks}

\correspondingauthor{Evan A. Rich}
\email{earich@umich.edu}

\author[0000-0002-1779-8181]{Evan A. Rich}
\affiliation{Department of Astronomy, University of Michigan, West Hall, 1085 South University Avenue, Ann Arbor, MI 48109-1090, USA}

\author[0000-0003-1534-5186]{Richard Teague}
\affiliation{Center for Astrophysics $|$ Harvard \& Smithsonian, 60 Garden Street, Cambridge, MA 02138, USA}

\author[0000-0002-3380-3307]{John D. Monnier}
\affiliation{Department of Astronomy, University of Michigan, West Hall, 1085 South University Avenue, Ann Arbor, MI 48109-1090, USA}

\author[0000-0001-9764-2357]{Claire L. Davies}
\affiliation{Astrophysics Group, University of Exeter, Stocker Road, Exeter, EX4 4QL, UK}

\author[0000-0003-4001-3589]{Arthur Bosman}
\affiliation{Department of Astronomy, University of Michigan, West Hall, 1085 South University Avenue, Ann Arbor, MI 48109-1090, USA}

\author[0000-0001-8228-9503]{Tim J. Harries}
\affiliation{Astrophysics Group, University of Exeter, Stocker Road, Exeter, EX4 4QL, UK}

\author[0000-0002-3950-5386]{Nuria Calvet}
\affiliation{Department of Astronomy, University of Michigan, West Hall, 1085 South University Avenue, Ann Arbor, MI 48109-1090, USA}

\author[0000-0002-8167-1767]{Fred C. Adams}
\affiliation{Department of Astronomy, University of Michigan, West Hall, 1085 South University Avenue, Ann Arbor, MI 48109-1090, USA}
\affiliation{Physics Department, University of Michigan, Randall Lab, 450 Church Street, Ann Arbor, MI 48109-1090, USA}

\author[0000-0003-1526-7587]{David Wilner}
\affiliation{Center for Astrophysics $|$ Harvard \& Smithsonian, 60 Garden Street, Cambridge, MA 02138, USA}

\author[0000-0003-3616-6822]{Zhaohuan Zhu}
\affiliation{University of Nevada, Las Vegas, NV, USA}

\begin{abstract}

Dust evolution in protoplanetary disks from small dust grains to pebbles is key to the planet formation process.
The gas in protoplanetary disks should influence the vertical distribution of small dust grains ($\sim$1 $\mu m$) in the disk.
Utilizing archival near-infrared polarized light and millimeter observations, we can measure the scale height and the flare parameter $\beta$ of the small dust grain scattering surface and $^{12}$CO gas emission surface for three protoplanetary disks IM Lup, HD 163296, and HD 97048 (CU Cha). For two systems, IM Lup and HD 163296, the $^{12}$CO gas and small dust grains at small radii from the star have similar heights but at larger radii ($>$100 au) the dust grain scattering surface height is lower than the $^{12}$CO gas emission surface height. In the case of HD 97048, the small dust grain scattering surface has similar heights to the $^{12}$CO gas emission surface at all radii. We ran a protoplanetary disk radiative transfer model of a generic protoplanetary disk with TORUS and showed that there is no difference between the observed scattering surface and $^{12}$CO emission surface. We also performed analytical modeling of the system and found that gas-to-dust ratios larger than 100 could explain the observed difference in IM Lup and HD 163296. This is the first direct comparison of observations of gas and small dust grain heights distribution in protoplanetary disks. Future observations of gas emission and near-infrared scattered light instruments are needed to look for similar trends in other protoplanetary disks.

\end{abstract}


\section{Introduction} \label{sec:intro}

Protoplanetary disks are composed of dust and gas and host forming and already formed exoplanets. Small dust grains ($\sim$ 1 $\mu m$) and gas interact due to the large amount of drag imparted on small dust grains through gas movement. Thus the vertical height of small dust grains in the disk should be related to the gas environment of the disk.
Thus, if we can independently measure the heights of gas and small dust grains in protoplanetary disks, we can probe fundamental disk characteristics such as the gas to dust ratio and turbulence in the disk.

It is expected that protoplanetary disks where the majority of the mass is in the central star, should have a flare shape. 
This is due to the fact that the internal temperature of the disk decreases slower than $r^{-1}$ \citep{k1987}. 
The flare shape has subsequently been confirmed through various direct observations of protoplanetary disks \citep{ginski2016, avenhaus2018,pinte2018,villenave2020}, and theoretically investigated \citep{meyer1982,fabien1991,bell1997,dalessio1998,birnstiel2010}. 

The radial height ($H(r)$) of protoplanetary disks is described by a powerlaw with $\beta$ as the flare parameter, $H_0$ as the scale height, and $r_0$ as the fiducial radius, i.e.,
\begin{equation}\label{eqn:disk_powerlaw}
    H(r) = H_0 \left( \frac{r}{r_0}\right)^{\beta} 
\end{equation}
We note that in this work we will use an $r_0$ = 100 au and all $H_0$ values have been adjusted to match this fiducial radius.  
Theoretical efforts have taken place to estimate the expected values $\beta$. \citet{chiang1997} show that for a disk with a surface density profile power law of $\sim$1.5 that is irradiated, the maximum flaring paramter value is $\beta$ = 9/7 $\sim$ 1.3. 
Similarly, \citet{k1987} shows that for a disk with a small disk to stellar mass ratio, the expected flaring parameter value should be $\beta = 9/8 = 1.125$ and derive a maximum flare parameter $\beta$ = 1.25. 

The disk flare parameter $\beta$ has previously been observationally measured for both gas and dust.
\citet{lagage2006} measured the height of polycyclic aromatic hydrocarbons (PAH) emission for HD 97048 measuring a radial minor axis offset resulting in a scale height $H_0$ of 34.2$^{+0.4}_{-2.2}$ au and $\beta$ = 1.26 $\pm$ 0.05. 
Additionally, the same technique can be used for multi-ringed systems for micron size particles in near-IR scattered light. \citet{ginski2016} observed HD 97048 with Very Large Telescope (VLT)/Spectro-Polarimetric High-contrast Exoplanet REsearch (SPHERE) and imaged the scattered light rings and gaps in the near-IR and measured a scale height $H_0$ of 18.5 au and a $\beta$ = 1.73 $\pm$ 0.05.
Finally, \citet{avenhaus2018} measured the $\beta$ flare parameter values in near-IR scattered light for V4046 Sgr (1.605 $\pm$ 0.132), RXJ 1615 (1.116 $\pm$ 0.095), and IM Lup (1.271 $\pm$ 0.197).
The same direct and model independent measurements can be preformed for $^{12}$CO gas emission utilizing the spatial and spectral sensitivity of Atacama Large Millimeter/submillimeter Array (ALMA). \citet{pinte2018} measured the $^{12}$CO and $^{13}$CO gas heights for IM Lup measuring $\beta$ values of 1.8 $\pm$ 0.2 and 2.1 $\pm$ 0.4 respectively. 
There have been other measurements of protoplanetary disk height (e.g. Class I source IRAS 04302+2247 \citealt{podio2020}) demonstrating that different molecules and Isotopologues probe different heights of the disk.
We note that these are all tracers of the gas and dust and are not direct measurements of the gas and dust height in the disk.

While both $^{12}$CO gas and small grain dust disk heights have been measured, the measurements have never been investigated or compared to test the fundamental properties of the gas vertically supporting small dust grains in protoplanetary disks.
In this paper, we will present small dust grain and $^{12}$CO gas height measurements in sections \ref{sec:small_measurements} and \ref{sec:CO_measurements}. 
We will then compare the CO gas and small grain dust height measurements in section \ref{sec:comparison}. 
We will then create a typical protoplanetary disk with radiative transfer modeling to test of the observed surfaces would bias the results (section \ref{sec:torus_model}). We then explore disk parameters that could potentially influence the gas and dust height with a simple analytical disk model in section \ref{sec:analytical_model}. 
Finally, we will discuss how these new findings will affect how we interpret protoplanetary disks and summarize our conclusions in section \ref{sec:conclusion}.

\section{Measurements of Gas and Small Dust Grains}

There are specific cases in which both the gas height of the disk as a function of radius and the small grain dust height can be measured. 
Protoplanetary disks must be moderately inclined ($\sim$30-70$^\circ$) as height measurement techniques for both take advantage of the three dimensional inclined disk projected onto the plane of the sky. 
Secondly, the CO gas must be plentiful and dense enough to create an optically thick disk photosphere that can be observed. 
Finally, the protoplanetary disk must host several scattered light rings in order to measure the small dust grain height as a function of radius. Such measurements probing the gas and dust heights has been made for a host of protoplanetary disks including HD 163296, IM Lup, HD 97048, RXJ 1615 \citep{ginski2016, avenhaus2018,monnier2017}.

To create our sample, we searched for previously observed protoplanetary systems that had a moderate inclination ($\sim$30-70$^{\circ}$), were young with plenty of CO gas, and were multi ringed to enable powerlaw fits of the small dust grains over multiple rings. We found that three targets: IM Lup, HD 163296 (MWC 275), and HD 97048 (CU Cha), have previous scattered light detection with multiple rings and sufficient ALMA observations to resolve the kinematic $^{12}$CO gas disk. Details about each of the three systems can be found in Table \ref{tbl:COgas}.
Below we will detail the previous detection's and the analysis needed to extract the scale heights ($H_0$) and flare parameter ($\beta$) for our target sample.

\begin{deluxetable*}{lccccccc}
\tablecaption{Gas Disk Parameters}
\tablehead{
\colhead{Object} & \colhead{Inc. ($^\circ$)} & \colhead{PA ($^\circ$)} &\colhead{Spectral Type} & \colhead{Age (Myr)} & \colhead{Distance (pc)} & \colhead{Mass (M\textsubscript{\(\odot\)})} & \colhead{Lum. (L\textsubscript{\(\odot\)})}}
\label{tbl:COgas}
\startdata
IM Lup & 48 $\pm$ 3 $^{(a)}$ & 143 $^{(a)}$ & M0 $^{(a)}$ & 1.1$\pm$0.2$^{(b)}$ & 155.8 $\pm$ 0.5 & 1 $^{(c)}$ & 0.9 $^{(d)}$ \\
HD 163296 & 42 $\pm$ 3$^{(e)}$ & 132$^{(e)}$ & A1$^{(f)}$ & 6.03$^{(g)}$ & 101.0 $\pm$ 0.4 & 1.95$^{(g)}$  & 20.4$^{(g)}$ \\
HD 97048 &  41 $\pm$ 3$^{(h)}$ & 2.8$^{(i)}$ & Be9.5/A0$^{(j)}$ & 2-3$^{(j)}$ & 184.4 $\pm$ 0.8 & 2.5$^{(j)}$ & 35$^{(k)}$ \\
\enddata
\tablecomments{The listed Inclination and disk major axis Position Angle (PA) are the assumed values for the CO extraction and are taken from the reference column. The assumed distances for these three targets are from EDR3 Gaia archive and is utilized throughout this work \citep{gaia2016,gaia2020}. Citations for specific values are from the following: (a) \citet{cleeves2016}, (b) \citet{avenhaus2018},
(c) \citet{panic2009}, (d) \citet{hughes1994}, (e) \citet{isella2016},
(f) \citet{Manoj2006}, (g) \citet{wichittanakom2020}, (h) \citet{walsh2016}, (i) \citet{ginski2016}, (j) \citet{ancker1998},
(k) \citet{vioque2018}.}
\end{deluxetable*}

\subsection{Small Dust Grain Height Measurements} \label{sec:small_measurements}

\begin{figure*}
    \centering
    \includegraphics[width=0.9\linewidth]{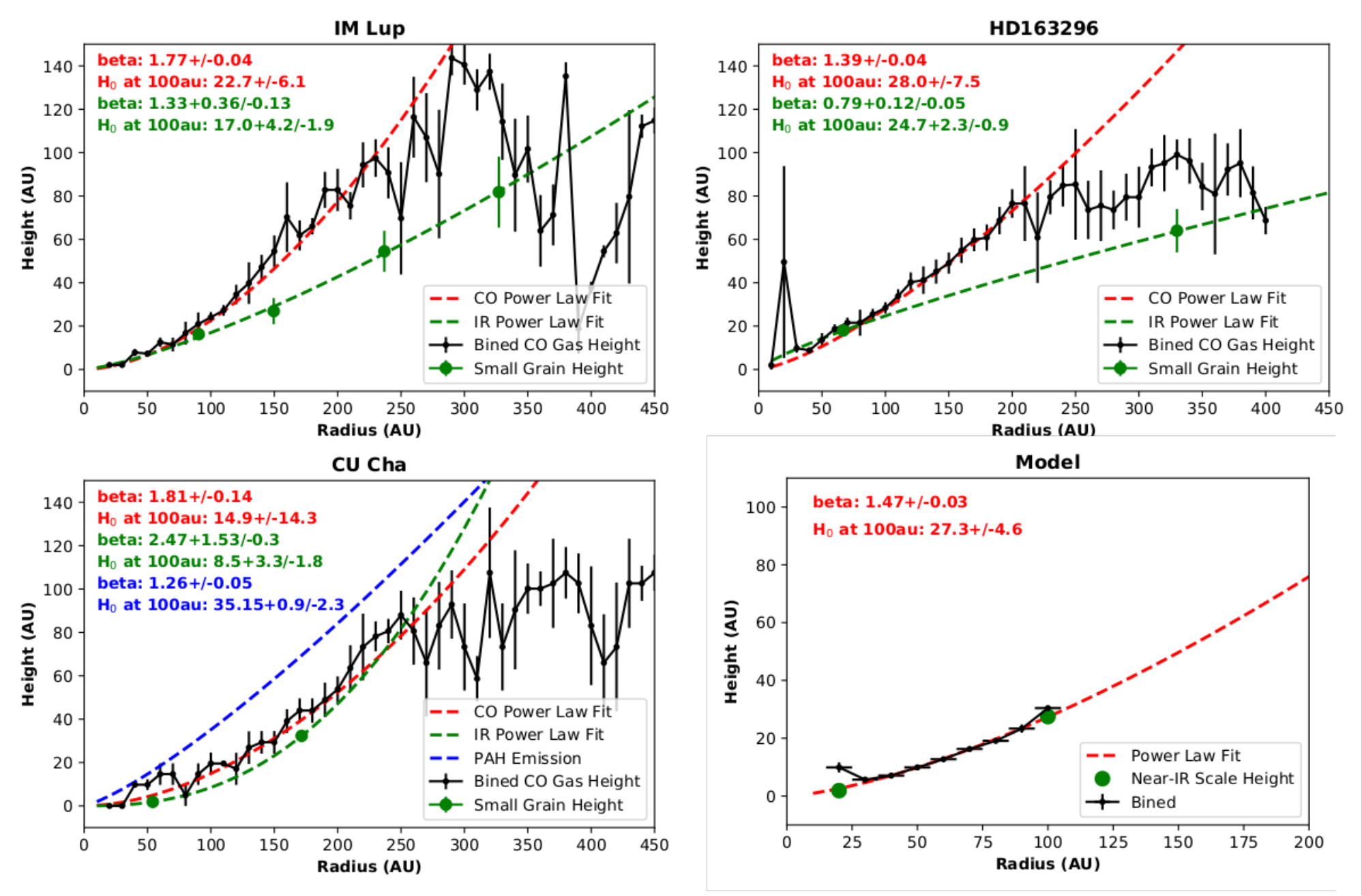}
    \caption{These figures show the measured heights of dust and gas as a function of radius for objects IM Lup (upper left), HD 163296 (upper right), HD 97048 (lower left), and a model protoplanetary disk (lower right) Measurements of the $^{12}$CO disk photosphere is shown in black, small dust grain height are shown in green, and PAH emission measured shown in blue published by \citet{lagage2006}. Power law fits values are show in the upper right hand corner of each plot which color corresponds to the color of the line. \label{fig:Scaled_heights}}
\end{figure*}

Scattered light imagery has shown that small dust grains accumulate in rings in protoplanetary disks.
If we assume that those rings are circular, we can fit ellipses to the peak flux then calculate the radius of the ring (major axis) along with a minor axis offset due to the height of the dust grains projected onto the plane of the sky. We will utilize previous literature measurements from \citet{avenhaus2018} (IM Lup), \citet{ginski2016} (HD 97048), and \citet{monnier2017} and \citet{rich2020} (HD 163296). The radii and measured heights of these rings can be found in Table \ref{tbl:small_grains} and the values are plotted in green in Figure ~\ref{fig:Scaled_heights}. We note that we re-interpret one of the 2nd ring height measurement made by \citet{ginski2016} which can be found in Appendix \ref{sec:CUrings}. With the exception of the second ring around HD 163296, all of the the other rings were imaged in polarimatry mode from instruments Gemini Planet Imager (GPI) or SPHERE. The 2nd ring listed from HD 163296 was taken in coronographic mode of the Hubble Space Telescope's Space Telescope Imager and Spectrograph (STIS) instrument. 
These observations are total intensity images and do not have any polarization information. We assume that the most common origin of the total intensity light from this second rind is from scattered light from the top of the disk and interpret the observations as we did for the other near-IR scattered light observations.
Additionally, STIS in coronographic mode is filter-less thus light from optical blue to near-IR was included in the image. The second ring in HD 163296 might be sensitive to smaller sized dust grains and at a different dust scattering height than the other rings imaged with either GPI or SPHERE in the near-IR.

HD 97048 and HD 163296 have further known rings that are not discussed in this work. In the case of HD 97048, there are two rings at a further radial extent to those listed in Table \ref{tbl:small_grains} \citep{ginski2016}. 
However, these two external rings have only been observed with total intensity imaging using Angular Differential Imaging reduction techniques which are known to alter the shape of continuous objects, thus we did not include these rings in our analysis. 
For HD 163296, there are two rings between the 77 au ring and the 360 au ring listed in Table \ref{tbl:small_grains}, but these have only been seen with ALMA in either mm dust grains or CO gas and never at optical or near-IR wavelengths  \citep{rich2020}.

We fit power-laws for each of the objects following equation \ref{eqn:disk_powerlaw} where the values of the $\beta$ flare parameter and scale height ($H_0$) are in Table \ref{tbl:flare_results} and plotted as the green dashed lines in Figure \ref{fig:Scaled_heights}. 
We used python's \textit{curve fit} package to fit the points and preformed a monte carlo of the point height uncertainties to estimate the uncertainty of $\beta$ and scale height ($H_0$). 
We explored a parameter space values of 0 to 4 for $\beta$ and 0 to 200 au for $H_0$. 
We note that the distribution of values for $\beta$ and $H_0$ are not symmetric around the median value, thus for the small dust grain results we have utilized asymmetric error bars as shown in Table \ref{tbl:flare_results}. For the HD 97048 result, the first ring has an error bar consistent with zero height from the mid-plane. 
This has resulted in $\beta$ values that can be un-physically large and the values presented should be considered lower-limits for the potential $\beta$ flare parameter values for small dust grains in HD 97048.

\begin{deluxetable*}{lccccc}
\tablecaption{Small grain ring heights}
\tablehead{
\colhead{Object} & \colhead{Ring} & \colhead{Radius (au)} & \colhead{Height (au)} & \colhead{Band} & \colhead{Reference}}
\label{tbl:small_grains}
\startdata
IM Lup & Ring 1 & 91.9 $\pm$ 3.17 & 16.5 $\pm$ 2.75 & H-band &  \citet{avenhaus2018} \\
IM Lup & Ring 2 & 152.11 $\pm$ 4.75 & 27.4 $\pm$ 6.1 & H-band & \citet{avenhaus2018} \\
IM Lup & Ring 3 & 240.84 $\pm$ 4.75 & 55.4 $\pm$ 9.6 & H-band & \citet{avenhaus2018}\\
IM Lup & Ring 4 & 332.75 $\pm$ 12.68 & 83.2 $\pm$ 16.6 & H-band & \citet{avenhaus2018} \\
HD 163296 & Ring 1 & 66.4$\pm$3.0 & 18.0 $\pm$ 2.0 & J-band & \citet{monnier2017} \\ 
HD 163296 & Ring 2 & 328$\pm$10.0 & 64.0 $\pm$ 10.0 & Clear Filter & \citet{rich2020} \\ 
HD 97048 & Ring 1 & 54.4$\pm$0.7 & 1.9$\pm$1.9 & J-band & \citet{ginski2016} \\
HD 97048 & Ring 2 & 172$\pm$4 & 32.3$\pm$1.5 & J-band & Appendix \ref{sec:CUrings} \\
\enddata
\tablecomments{This table lists the ring heights and mid-plane radial distances of each of the objects. The assumed distances to each target are from Gaia EDR3 and can be found in Table \ref{tbl:COgas}.}
\end{deluxetable*}

\subsection{CO Gas Height Measurements} \label{sec:CO_measurements}

Data for the $^{12}$CO gas measurements are from previous works or utilize the ALMA archive reductions. Observations of HD 163296 and IM Lup were taken as part of the Disk Substructure at High Angular Resolution (DSHARP) program (Program ID 2016.1.00484.L) \citep{andrews2018}\footnote{https://almascience.eso.org/almadata/lp/DSHARP/}. 
The DSHARP archive provides calibrated $^{12}$CO data cubes.  See \citet{andrews2018} for a full description of the data reduction description. 
HD 97048 data was taken from the ALMA archive (Program ID 2016.1.00826.S) reduced using the ALMA automatic reduction pipeline which produced the $^{12}$CO data cube. 

To measure the height of the CO gas disk, we utilize the techniques first described in \citet{pinte2018}, and use a the python package by \citet{teague2019}. In summary, the program rotates the CO disk cube such that the disk major axis PA in the data cube is horizontal and that the closest side of the of the disk to Earth is at the top of the image. Pixels in data cube with low signal to noise are masked prior to finding the maximum flux values. 
Vertical pixel slices are taken along the disk in every frame in the cube to find local maxima corresponding to the $^{12}$CO emission. The brightest flux pixel of each column are taken to be the front side of the disk and the second brightest flux pixel is the bottom of the disk. Using the pixel locations of the front ($x$,$y_f$) and back ($x$,$y_b$), sides of the disk, with the prior knowledge of the disks inclination ($i$), and the stars location ($x_s$, $y_s$), the mid-plane radius $r$ is given by 
\begin{equation}\label{eqn:radius}
    r = \sqrt{(x-x_s)^2 + \left(\frac{y_f - y_c}{\cos(i)} \right)^2},
\end{equation}
and height $h$ has the form
\begin{equation}\label{eqn:height}
    h = \frac{y_c-y_s}{\sin(i)}.
\end{equation}
The y coordinate center ($y_c$) is the average of the $y_f$ and $y_b$ values.

The python package outputs a set of radii ($r$), heights ($h$), and flux values for the disk. 
We further refined the set of points by removing any heights that had negative values and radii that were beyond 4". 
We binned the data in mid-plane radius with bin sizes of 10 au and applied a median clip to remove further extraneous points. 
The binned median values along with error bars estimated via median absolute deviation (MAD) and the inclination uncertainty added in quadrature are shown in Figure \ref{fig:Scaled_heights}.

We fit the power-law in log-log space with the best fit values shown in Table \ref{tbl:flare_results} and plotted in Figure \ref{fig:Scaled_heights}. The fits were done on the individual radial and height points and the linear regression fits were bootstrapped to estimate the errors. All three targets have a broken power law distribution, thus we only fit between 50 and 250 au.

\begin{deluxetable*}{lcccc}[h]
\tablecaption{Gas and small grain power-law fit results}
\tablehead{
\colhead{Object} & \colhead{$^{12}$CO $\beta$} & \colhead{$^{12}$CO $H_0$ (au)} & \colhead{Dust $\beta$} & \colhead{Dust $H_0$ (au)} \\
\colhead{} & \colhead{} & \colhead{at r=100au} & \colhead{} & \colhead{at r=100au}}
\label{tbl:flare_results}
\startdata
IM Lup & 1.77 $\pm$ 0.04 & 22.7 $\pm$ 6.1 & 1.34$^{+0.35}_{-0.13}$ & 17.0$^{+4.1}_{-1.8}$ \\
HD 163296 & 1.39 $\pm$ 0.04 & 28.0 $\pm$ 7.5 & 0.79$^{+0.12}_{-0.05}$ & 24.7$^{+2.3}_{-0.9}$  \\
HD 97048 & 1.81 $\pm$ 0.14 & 14.9 $\pm$ 14.3 & 2.48$^{+1.52}_{-0.3}$ & 8.5$^{+3.3}_{-1.8}$ \\
\enddata
\tablecomments{This table lists measured $\beta$ flare parameter and $H_0$ scale height for small dust grains and $^{12}$CO plotted in Figure \ref{fig:Scaled_heights}. 
Parameters $H_0$ and $\beta$ are defined in Equation \ref{eqn:disk_powerlaw}.}
\end{deluxetable*}

\section{Comparison of CO gas and Small Dust Grains} \label{sec:comparison}

We were able to measure the $^{12}$CO gas flare parameter $\beta$ for IM Lup (1.77$\pm$0.04), HD 163296 (1.39$\pm$0.04), and HD 97048 (1.81$\pm$0.14). Our IM Lup $\beta$ measurement is well matched to the previous measurement of 1.8$\pm$0.2 \citep{pinte2018}, and is the only CO gas flare parameter $\beta$ that had previously been measured. Both IM Lup and HD 97048 both have $\beta$ parameter values that are much larger than the estimated maximum theoretical values (1.3 \citealt{chiang1997};, 1.25 \citealt{k1987}). However, $\beta$ parameter of HD 163296 is more consistent with the theoretical maximum values. This could be an indication of age as HD 163296 is estimated to be older than IM Lup and HD 97048 (See Table \ref{tbl:COgas}).

We were also able to measure the small grain dust $\beta$ flare parameter for IM Lup (1.34$^{+0.35}_{-0.13}$), HD 163296 (0.79$^{+0.12}_{-0.05}$) and HD 97048 (2.48$^{+1.52}_{-0.3}$). 
Both IM Lup and HD 163296 have $\beta$ flare parameter values that are broadly consistent with a maximum theoretical $\beta$ value (1.3 \citealt{chiang1997};, 1.25 \citealt{k1987}), while HD 97048's measured $\beta$ is much larger. This was previously noted by \citet{avenhaus2018, pinte2018}
We do not have a good constraint on the $\beta$ parameter value for HD 163296 and HD 97048 as there are only two points. 
Our measured values for IM Lup are consistent with the values measured by \citet{avenhaus2018} (1.271 $\pm$ 0.197), while our HD 97048 measurement is much larger than \citet{ginski2016} (1.73 $\pm$ 0.05). 
However, \citet{ginski2016} utilized height measurements for all four rings and the heights of the gaps and we have re-calculated the estimated height for the second ring in HD 97048 in Appendix \ref{sec:CUrings}. 
Thus our estimate is more conservative as we avoid ADI total intensity artifacts as discussed above and the assumption of a large wall at ring 2 in the HD 97048 system.

Here we caution against direct comparisons between the observed emission and scattering surface of the $^{12}$CO gas and small dust grain surfaces to theoretical values of the gas and dust heights, as previously noted by \citet{avenhaus2018,pinte2018}. As noted in section \ref{sec:intro}, the measured flaring parameter $\beta$ is not actually physical scale heights of the disk but the observed scale heights that are dependent on the temperature of the disk (for CO) and the dust scattering properties of the small grain dust. We will begin to address this issue in section \ref{sec:torus_model}.

Finally, we compare the height as a function of radius of CO gas and the small dust grains. 
Notably, in all three systems the first ring is co-located with the $^{12}$CO gas emisson layer as shown in Figure \ref{fig:Scaled_heights}. For systems IM Lup and HD 163296, the scattered light rings have smaller heights than the $^{12}$CO gas emission layer. 
This is confirmed as the $^{12}$CO gas has a $\beta$ flare parameter value of 1.77$\pm$0.04 that is 1-sigma larger than the small dust grains $\beta$ of 1.34$^{0.35}_{-0.13}$. 
Though there are fewer rings, HD 163296 seems to show the same trend as IM Lup. 
However, this same trend is not seen in the case of HD 97048, where both of the two rings appear to be co-located with the CO gas and the $\beta$ flare parameter are larger rather than smaller. 
Thus, in the case of some protoplanetary disks, it appears that there is a radial dependence of a disk when comparing the $^{12}$CO gas emission height and the small dust grains height. We note that the second ring in HD 163296 and the forth ring of IM Lup are located exterior to $^{12}$CO power-law break.

\section{Simple Protoplanetary Disk Model} \label{sec:torus_model}

An issue remains in that we are not tracing the true gas or dust scale height of the disk. In the case of the $^{12}$CO gas, the emission comes from the photosphere of the disk which is dependent on the local temperature of the disk and the $^{12}$CO abundance. For the small dust grains, the height is tracing the local illumination from the protostar onto the disk, and the scattering efficiency of the dust grains. In both cases, and especially the scattering efficiency of the small dust grains, to calculate the true gas or dust scale height from the observables is difficult due to degeneracy's and unknown quantities of the disk material. We choose to approach this problem by preforming radiative transfer modeling of a generic two ringed protoplanetary disks and see if a model $^{12}$ CO ALMA image and an H-band scattered light image show any relative difference.

We model our simple protoplanetary disk using the radiative transfer program TORUS \citep{harries2000,harries2004,rundle2010,harries2011}. 
TORUS is a Monte Carlo radiative transfer code using an adaptive mesh and radiative equilibrium method described in \citet{lucy1999}. We utilized atomic and molecular lines from the LAMDA database \citep{schoier2005}, and CO energy level coefficients from \citet{muller2005}.
We parameterized a generic protoplanetary disk in 2D to be composted of CO gas, small dust grains 0.7 - 2 $\mu m$, and large dust grains at 0.9-1 mm using a grain prescription from \citet{draine1984}. Stellar and disk parameters are based on  \citet{rich2019} which modeled the protoplanetary disk system HD 163296.
The disk has a disk density exponent $\alpha$ of 2.0 and a disk scale height exponent $\beta$ of 1.4 for the dust and gas. 
The dust grains are defined as a fraction of the gas height with the small dust grains having 70\% of the gas scale height and the large dust grains having 10\% of the gas scale height. 
We modeled two rings at 25 and 100 au from the star with ring widths of 1 au. We included CO freeze out for $T < 30$K and CO dissociation using quenching rates from \citet{yang2010}. This should accurately replicate the $^{12}$CO disk emission surface observed by ALMA. A subset of model parameters is shown in Table \ref{tbl:Model_Paramters}.

\begin{deluxetable}{lc}
\tablecaption{Model Parameters}
\tablehead{
\colhead{Parameter} & \colhead{Value}}
\label{tbl:Model_Paramters}
\startdata
T$_{\textrm{eff}}$ (K) & 9250 \\
M$_{\textrm{star}}$ (M\textsubscript{\(\odot\)}) & 2.5 \\
M$_{\textrm{disk}}$ (M\textsubscript{\(\odot\)}) & 0.2 \\
disk density exponent ($\alpha_{den}$) & 2.0 \\
disk scale height exponent ($\beta$) & 1.4 \\
disk scale height (H$_0$; au) & 10 \\
small dust grains size range ($\mu$m) & 0.7--2 \\
large dust grains size range (mm) & 0.9--1 \\
\enddata
\tablecomments{A subset of TORUS parameters for the simple protoplanetary disk model. $^{12}$CO emission surface and small dust grain scattering surface measurements are shown in Figure \ref{fig:Scaled_heights}. Note that the disk density exponent $\alpha$ is different from the $\alpha$ turbulent viscosity coefficient discussed in section \ref{sec:analytical_model}.}
\end{deluxetable}

Using the model parameters described above, TORUS produced a scattered light H-band image of the two scattered light rings and a $^{12}$CO data cube. The two model data sets were convolved to match SPHERE/IRDIS and ALMA resolution respectively. We processed the model $^{12}$CO data cubes using the same process described in Section \ref{sec:CO_measurements}. Similarly, we fit ellipses to the rings as described in the Appendix \ref{sec:CUrings}. The results are plotted in Figure \ref{fig:Scaled_heights} with the green dots the small grain heights, the black points are the CO gas height, and the red dotted line is the best fit power-law. 
As shown in Figure \ref{fig:Scaled_heights}, both the of the scattered light rings are co-located with the $^{12}$CO gas emission. The measured flare parameter $\beta$ is 1.47 $\pm$ 0.03, very similar to the input disk scale height exponent of 1.4. Thus we can see the effect that the flare shape of the $^{12}$CO gas is not necessarily the same as the true CO gas flare shape. Finally, our simple model demonstrates that our observable tracers of the dust and gas height, $^{12}$CO emission surface and the small dust grain scattering surface, have similar flare profiles.

\section{Analytical Disk Model} \label{sec:analytical_model}

We next want to explore the disk parameters that could cause the difference in $\beta$ flare parameter values between the small dust grains scattering surface and $^{12}$CO gas emission. 
We choose to use a parametric structure model of an exponentially tampered accretion disk profile in hydrostatic equilibrium and use the parameterizations and equations outlined \citet{williams2014}. 
In summary, We define the pressure scale height, $H_P$, as
\begin{equation}\label{eqn:pressure_height}
    H_P = \frac{kT_{\textrm{mid}}r^3}{GM_{\textrm{star}}\mu m_H},
\end{equation}
where $r$ is the mid-plane radius of the disk, $T_{\textrm{mid}}$ is the mid-plane temperature of the disk at $r$, $M_{\textrm{star}}$ is the mass of the protostar, and $\mu$ is the mean molecular weight of the gas. The mid-plane temperature is defined as a powerlaw as 
\begin{equation}\label{eqn:T_mid}
   T_{\textrm{mid}}(r) = T_{\textrm{mid,1}} \left( \frac{r}{1 AU} \right) ^{-q},
\end{equation}
where we assume values of $T_{\textrm{mid,1}}$ = 200 K and $q$ = 0.55 as assumed in \citet{williams2014}.

\begin{figure*}[h]
    \centering
    \includegraphics[width=0.99\linewidth]{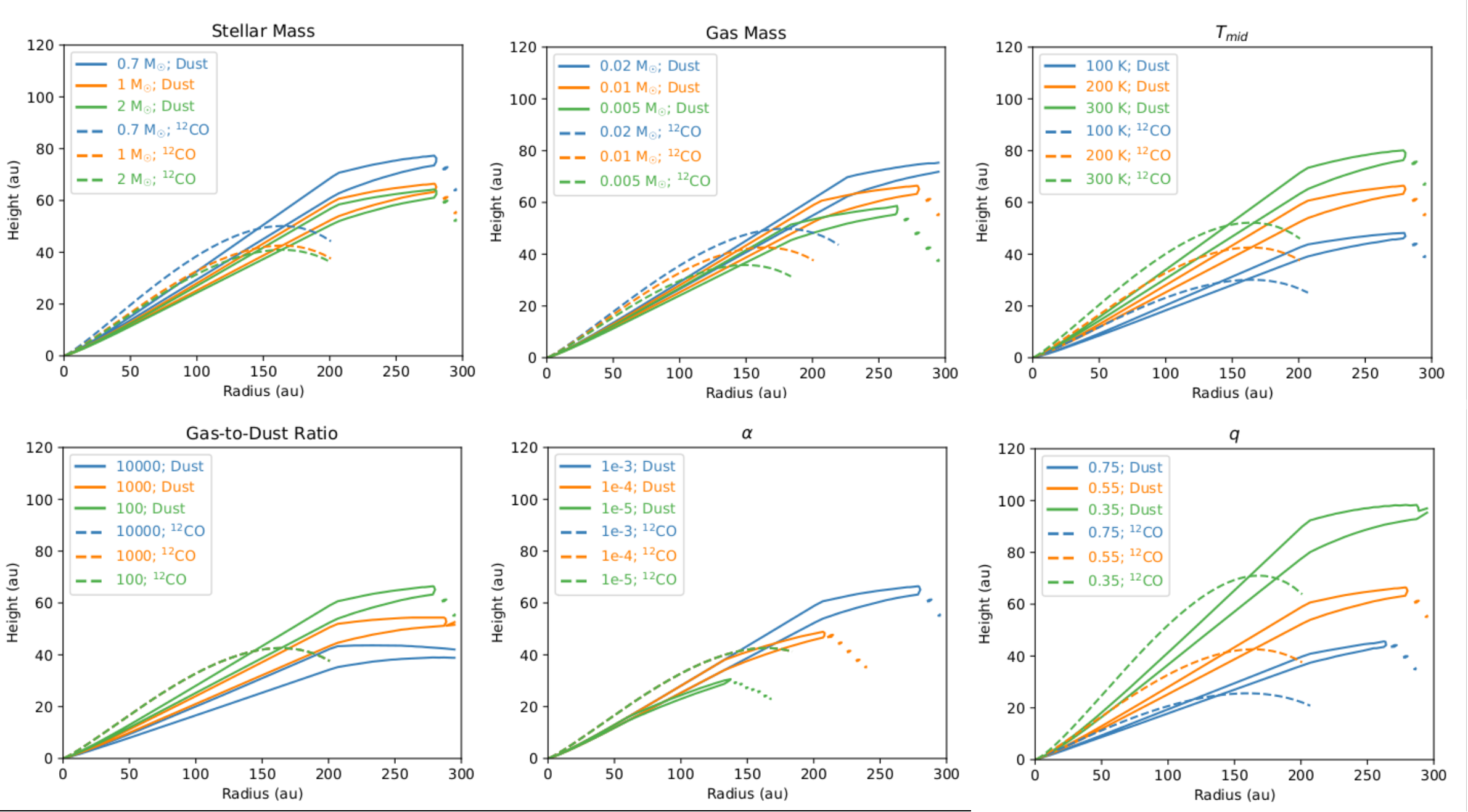}
    \caption{These six panels show the results of analytical modeling varying the Solar Mass (top left), Disk Gas Mass (top middle), T$_{mid}$ (top right), Gas-to-Dust Ratio (bottom left), $\alpha$ (bottom middle), and mid-plane temperature exponent $q$ (bottom right) parameters. The dashed lines show the upper region of the CO emitting photosphere of the disk, and the solid line colored contours show the scattering surface containing 80$\%$ of the scattering surface. 
    The base model are defined by the parameters: $\gamma$ = 0.75, $\psi$ = 0.2, R$_c$ = 60 au, R$_{\textrm{in}}$ = 1.0 au, M$_{\textrm{gas}}$ = 0.01M\textsubscript{\(\odot\)}, M$_{\textrm{star}}$ = 1M\textsubscript{\(\odot\)}, T$_{\textrm{mid}}$ = 200K, T$_{\textrm{atm}}$ = 1000 K, $\alpha=10^{-3}$, q=0.55, a=1e-5.    \label{fig:parameters}}
\end{figure*}

The temperature structure of the disk, utilized in \citet{williams2014}, defines $^{12}$CO emission region. 
For temperatures $T< 20$K, the CO will freeze-out creating the lower boundary region, and the upper boundary is defined by CO dissociation when the column density of H$_2$ reaches 1.3$\times$10$^{21}$ H$_2$ cm$^{-2}$. 
Since the $^{12}$CO emission layer is optically thick, the location of the $^{12}$CO emission will be slightly below the upper boundary, thus we only plot the upper $^{12}$CO emission as a tracer for the $^{12}CO$ emission. 
The $^{12}$CO upper emission layers are shown in Figure \ref{fig:parameters} as dashed lines.

Next, we can calculate the small dust grain pressure height using the gas pressure height. We follow the parameterization of dust pressure height ($H_d$) as outlined in \citet{pohl2017} as 
\begin{equation}\label{eqn:dust_pressure}
   H_d (r,a) = H_p (r) \times min \left(1, \sqrt{\frac{\alpha}{min(\textrm{St}, 1/2)(1 + \textrm{St}^2}} \right).
\end{equation}
We define $\alpha$ as the viscosity parameter, a as the dust grain diameter, r as the mid-plane radius, and St as the Stokes number where St = $\rho$a$\pi$/$\Sigma_g$2. We utilize the same gas surface density power-law with exponential taper ($\Sigma_g$) as defined in \citet{williams2014}, and assume a dust grain volume density of 1.2 g cm$^{-3}$ as used in \citet{pohl2017}. 

Having obtained the dust pressure height (H$_d$), we now estimate where the dust scattering layer is located. Using the dust opacities, we calculated the line of sight from the star to that radial bin in the disk and found the height at which the largest number of scattering events would occur. We plotted contours that are 80\% of the maximum scattering events that occur as shown in Figure \ref{fig:parameters} as colored solid lines. Thus, the values plotted in Figure \ref{fig:parameters} are parametric models of the observed quantities plotted in Figure \ref{fig:Scaled_heights}.

We explore the effect of six parameters (M$_{\textrm{star}}$, M$_{\textrm{gas}}$, T$_mid$, Gas to Dust Ratio, viscosity parameter $\alpha$, $q$) have on the difference between the $^{12}$CO emission  and the small dust grain scattering surface. 
The first parameter, M$_{\textrm{star}}$, has a direct effect the pressure scale height of the gas as shown in equation \ref{eqn:pressure_height}. 
We note that with changing stellar mass, we also expect a change in stellar luminosity ($L$) thus a change in the mid-plane temperature of the disk ($T_{\textrm{mid}}$) which will effect the disk height structure as well. 
In order to account for this, we utilize equation 12c in \citet{chiang1997} where the internal disk temperature $T_{\textrm{mid}} \propto $ $T_{\textrm{eff}} \times R_{\textrm{*}}^{1/2}$. 
Thus, the luminosity of the star $L \propto T_{\textrm{mid}}^4$. 
As seen in Equation \ref{eqn:pressure_height}, $T_{\textrm{mid}}$ and $M_{\textrm{star}}$ terms are proportional thus any change in $M_{\textrm{star}}$ could be cancelled out by an equal change in $T_{\textrm{mid}}$. However, the mass and luminosity relationship for protostars is complicated, thus we assumed mass and luminosities for two of our objects, IM Lup (0.7 M\textsubscript{\(\odot\)}, 1.56 L\textsubscript{\(\odot\)}) and HD 163296 (1.95 M\textsubscript{\(\odot\)}, 20.4 L\textsubscript{\(\odot\)}) \citep{avenhaus2018,wichittanakom2020}. 
We scaled the $T_{\textrm{mid}}$ value from the luminosity assuming a luminosity of 1.56, 1.75, and 20.4 for each of the corresponding solar mass values of 0.7, 1.0, and 2 M\textsubscript{\(\odot\)}. As seen in Figure \ref{fig:parameters}, the the $^{12}$CO emission height and small dust grain scattering surface are equally affected by the change of stellar mass and luminosity. 

Next we varied the total gas mass M$_{\textrm{gas}}$ of the system which changes the surface density of the gas. As shown in the top middle subplot in Figure \ref{fig:parameters}, this parameter equally effects both the $^{12}$CO gas emission height and small dust grain scattering surface height. Thus the total gas mass M$_{\textrm{gas}}$ cannot explain the difference in the observed gas and dust heights in these disks.

We explored the global Gas-to-Dust ratio in the disk. We assumed nominal values of G/D=100 and explored parameter values by factors of 10 to see if these values could explain the gas and dust height differences. As shown in Figure \ref{fig:parameters}, changing the gas to dust ratio has no effect on the observed $^{12}$CO gas height (dashed lines are plotted on-top of each other) but do effect the small dust grain scattering surface heights. By varying the gas-to-dust ratio we are effectively lowering the amount of dust in the system thus lowering the height of the scattering layer of the disk. This appears to replicate what we observe in the IM Lup and HD 163296 system as shown in Figure \ref{fig:Scaled_heights}. We note that we are assuming in our simplified models is that the gas to dust ratio is constant throughout the disk. This is unlikely to be true both in terms of height from the mid-plane due to dust settling. More sophisticated modeling is needed with non-constant gas to dust ratio to show that the similar trend is true.

We explored the effects of the assumed disk temperature structure might have on the radial height of $^{12}CO$ and small dust grains by varying the mid-plane temperature, T$_{mid}$, and the mid-plane temperature exponent $q$ as defined in Eqn \ref{eqn:T_mid}. We note that changing T$_{mid}$ is similar to changing the luminosity of the star as discussed above when varying the stellar mass parameter. When decreasing values of T$_{mid}$, the slope of both the $^{12}CO$ and small dust grain surfaces decreases, with the slope of the $^{12}CO$ surface decreasing faster. However, this is unlikely to explain the surface height differences as IM Lup and CU Cha have similar $\beta$ values but different resulting surface heights for small dust grains and $^{12}CO$ gas. Next, we varied $q$ which had an inverse effect on $^{12}CO$ and small dust grain surfaces as T$_{mid}$ parameter did, with increasing the value of $q$, the slopes of $^{12}CO$ and small dust grain surfaces were smaller. Similarly, the $^{12}CO$ surface slope decreased faster. However, in a similar argument for T$_{mid}$, this would not explain the difference between IM Lup and CU Cha.

The last parameter explored was the $\alpha$ turbulent viscosity coefficient shown in Figure \ref{fig:parameters}. Note that the turbulent visocisty coefficient is usually denoted as $\alpha$ in the literature, which is not to be confused with the disk density exponent $\alpha$ used in section \ref{sec:torus_model}.  
We assumed a nominal value of $\alpha=10^{-3}$ and explored factors of 10 smaller. Similar to the gas to dust ratio, the $\alpha$ viscosity parameter does not effect the $^{12}$CO gas emission height.
By decreasing the value of the $\alpha$ viscosity coefficient, the scattering surface for small dust grains is lower. 
The $\alpha$ viscosity parameter could potentially explain the scale height deviation we observe for HD 163296 and IM Lup as shown in Figure \ref{fig:Scaled_heights}. 
However, the amount of turbulence in IM Lup and HD 163296 have been measured roughly translating to an $\alpha$ = 1$\times$10$^{-3}$ \citep{hughes2011,flaherty2017}. 
A lower value of $\alpha$ is needed to explain the height disparity in small grains versus $^{12}$CO gas by a factor of $\times$10-100 making this explanation unlikely.

\section{Discussion and Conclusions} \label{sec:conclusion}

We have measured the flaring parameter $\beta$ for $^{12}$CO gas emission surface of protoplanetary disks around IM Lup, HD 163296, and HD 97048 and compared them to previous measurements of small dust grain scattering surface heights. 
We find that for IM Lup and HD 163296, small dust grain heights are co-located with the CO gas emission at small radii from the star, but not at radii larger than $>$ 100 au. 
However, for HD 97048 we find that the small dust grain heights and the CO gas emission are co-located throughout the disk. 
With simple radiative transfer modeling of a protoplanetary disk, we were able to show the expectation that the small dust grain scattering surface is at a similar height to the $^{12}$CO gas emission layer.

Though we cannot definitively determine any trends with a sample size of 3 disks, we can look for potential trends for future investigations. 
While the disks around IM Lup and HD 163296 show the same radial trend, the two systems have little in common as HD 163296 (7.6 Myr) is much older than IM Lup (1.1 Myr), IM Lup is a T-Tauri M0 star while HD 163296 is a Herbig-Ae A1, and IM Lup is a solar mass star (1 M\textsubscript{\(\odot\)}) and HD 163296 is twice as massive (1.95 M\textsubscript{\(\odot\)}). 
In fact, HD 163296 has the most in common with HD 97048 as they are both Herbig stars and are both more massive with HD 97048 being the most massive star at (2.5 M\textsubscript{\(\odot\)}).
It appears that vertical heights of gas and dust trends may not simply be a function of mass, age, or spectral type.
Increasing the number of protoplanetary disks in which we have measured gas and small dust grain heights is necessary to fully investigate disk parameter trends such as age, mass, and stellar host star.

There are several caveats when comparing our sample of three targets to a generic protoplanetary disk. 
First, the mechanism(s) for formation of the dust rings themselves are unknown (e.g. protoplanets, ice lines), and these mechanisms and environments could potentially influence the vertical distribution of the small dust grains within these rings.
For example, Illumination from the star or from an accreting protoplanet can increase the vertical height of the disk similar to what occurs in the inner disk \citep{natta2001,Dullemond2001,Montesinos2021}. 
If these mechanisms play an important role in the vertical distribution of small dust grains, what we are observing may not be ubiquitous for all protoplanetary disks but for only protoplanetary disks that have rings caused by the mechanism at hand. Modeling of the various mechanisms that cause ring formation in protoplanetary disks and the influence of the vertical gas and dust structure is needed. 
Secondly, we assume that these rings are circular and centered around the central star, however there is evidence in some systems (eg. GW Ori: \citealt{kraus2020}, HD169142: \citealt{bertrang2018}) have off-set rings thus this assumption is defiantly not valid for all protoplanetary disk systems.
Alternative measurements of small dust grain heights that are not dependent on the presence of rings in protoplanetary disks while also dealing with the intricacies of interpreting scattered light images are needed to avoid the dust ring issues. Finally, different isotopologues of CO and different molecules will trace lower layers in the disk that might correlate differently with the small dust grain heights than $^{12}$CO. Future work is needed to investigate the correlation of other molecule emission heights in the disk to small dust grain scattering surface heights.

One issue of our sample is that these systems have much larger disks than a typical protoplanetary disk of around 100 au. What we could be observing is that beyond 100 au, the disks are not behaving like our archetypal model as described in our analytical model (section \ref{sec:analytical_model}). Future investigations of more typically sized protoplanetary disks is necessary to verify if the observed dust and gas height deviation is present in smaller disks as well.
In line with this caveat, we also assume that our disks have a continuous power-law with no breaks at all. However, power-law breaks have previously been identified in protoplanetary disk systems for both scattered light and $^{12}$CO gas \citep{wisniewski2008,pinte2018,teague2018}. This aspect is important as two scattered light rings, in HD 163296 and IM Lup, are exterior to the $^{12}$CO gas power-law break. Additionally, the optical depth of the $^{12}$CO gas could play an important part in the outer disk as the gas becomes optically thin. While we are unable to investigate this caveat due to the limited number of rings in our systems, future work needs to investigate the effect of scale height with broken power-law height systems.

We investigated six stellar and disk parameters (M$_{\textrm{star}}$, M$_{\textrm{gas}}$, T$_{mid}$, Gas-to-dust ratio, $\alpha$, $q$) that might explain the flared discrepancy of gas and dust as seen in IM Lup and HD 163296. The stellar mass, total gas mass, T$_{mid}$, and $q$ parameters are unlikely to explain the discrepant flare trend. Both the gas-to-dust ratio, and viscosity parameter $\alpha$ could potentially explain the $\beta$ flare parameter in IM Lup and HD 163296. However, the $\alpha$ viscosity parameter would have to be unusually small and requires further investigation. If the Gas-to-dust ratio is largely responsible to explain the discrepancy between the difference in $\beta$ flare parameters for small dust grains and gas, our height comparison  methods might be good tracers of the gas-to-dust ratios in the top layers of protoplanetary disks. We note that this treatment of disk temperature is very simplistic and future research on its effect on the height of small dust grains and $^12$CO gas is needed.

Another potential mechanism to explain the radial dependence of the CO gas and small dust grain height deviation is settling. Modeling by \citet{facchini2017} shows that lower turbulence in the disk can result the same type of radial dependence we observe where the small dust grains appear to decouple in height from the CO gas as you travel outwards radially. Thus, by measuring the CO gas and the small dust grains, we could potentially constrain the turbulence in the disk. 
This aligns with our analytical model discussed in section \ref{sec:analytical_model}. 
However, there have been turbulence measurements of HD 163296 which are consistent with an viscosity parameter value of $\alpha$ = 1$\times$10$^{-3}$ \citep{hughes2011,flaherty2017}. 
According to our analytical model in section \ref{sec:analytical_model}, this amount of viscosity in the disk may not be enough to explain the CO gas and the small dust grain heights. Thus, lower turbulence in the disk is unlikely to have caused the radial dependence of the CO gas and small dust grain height deviation for HD 163296.
More observationally direct measurements of the protoplantary disk turbulence to investigate is HD 163296 is an outlier and turbulence measurements of HD 97048 and IM Lup for comparison.
Future hydro-dynamical modeling work is needed that includes an exploration of the turbulence in the disk that also outputs gas and dust scale height observables such as the $^{12}$CO emission surface and the small dust grain scattering surface. 

Thanks to Dr. Henning Avenhaus and Dr. Christian Ginski for supplying SPHERE/IRDIS images for IM Lup and HD 97048 respectively. E.A.R acknowledge support from NSF AST 1830728. This work has made use of data from the European Space Agency (ESA) mission {\it Gaia} (\url{https://www.cosmos.esa.int/gaia}), processed by the {\it Gaia}
Data Processing and Analysis Consortium (DPAC, \url{https://www.cosmos.esa.int/web/gaia/dpac/consortium}). Funding for the DPAC has been provided by national institutions, in particular the institutions participating in the {\it Gaia} Multilateral Agreement.

\appendix
\section{Re-interpreting the Rings around HD 97048}\label{sec:CUrings}

\begin{figure}[h]
    \centering
    \includegraphics[width=0.99\linewidth]{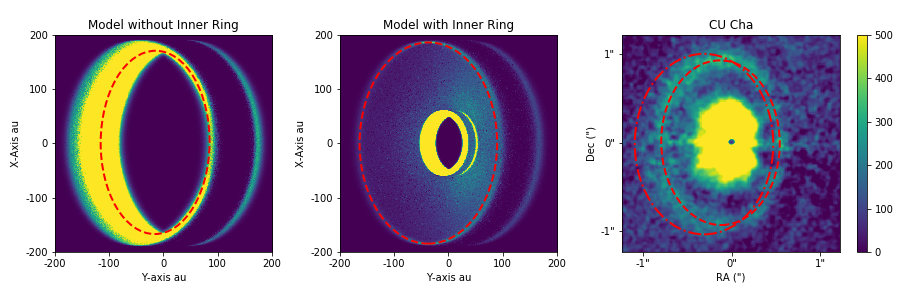}
    \caption{This Figure shows polarized light Torus models (left and center) demonstrate that the second ring in the protoplanetary disk HD 97048 (right) is unlikely to have a large wall. The left panel is a Torus model with a dusty ring located at 188 au and the dashed red line shows the best ellipse fit of the disk. The middle panel is a Torus model with two dusty rings located at 55 au and 188 au, and the red dashed line is the best fit ellipse for the outer ring. The right panel shows the scattered light J-band image of HD 97048 with the red dashed line as our best fit ellipse to the observed data and the dot-dashed line as the best fit ellipse from \citet{ginski2016}. \label{fig:no_innerring}}
\end{figure}

Previous near-IR observations with VLT/SPHERE of the Protoplanetary disk HD 97048 (CU Cha) by \citet{ginski2016} found four dusty rings at 46 au, 161 au, 272 au, and 341 au. \citet{ginski2016} fit ellipses to the rings where the first two inner ring fits (46 and 161 au) were to scattered light imaging and the two outer ring fits (272 and 341 au) were to total intensity imaging with a Angular Differential Imaging techniques reduction. For all four rings, \citet{ginski2016} measured a minor axis offset associated with a projected scale height of the disk. They note that the second ring might possibly be biased by hosting a partially illuminated``wall'' which would influce the measured height on the far side of the disk (left side). Thus, the flux observed is not only from the top of the ring, but closer to the midplane of the disk. This can bias the ellipse fitting of the ring as the wall will be fully illuminated from the view of the observer making the minor axis of the ring appear to be smaller than in reality (See left panel of Figure \ref{fig:no_innerring}). To avoid the ring location being biased by the wall, \citet{ginski2016} only fit an ellipse to the forward part of the disk, avoiding areas where a potential wall is fully illuminated. The explanation of a ``walls'' being present at the 2nd ring is unexpected as ``walls'' are typically associated with the inner disk, and in the case of HD 97048 there is scattered light flux traced to the inner working angle of the VLT/SPHERE images. We would expect that the inner small dust grain material would shadow the 2nd ring and not allow for the wall to be illuminated.

To test whether you expect to be strongly biased by a wall in HD 97048 with the presence of inner disk material, we preformed a simple scattered light radiative transfer modeling using TORUS of a single ringed system and a two ringed disk shown in Figure \ref{fig:no_innerring}. We utilized the same TORUS model parameters as described in section \ref{sec:torus_model} with the exception that the two rings are located at 46 au and 188 au. The left panel in \ref{fig:no_innerring} shows a single ring located at 188 au from the star and is fully illuminated creating a wall. The dashed red line shows the best fit ellipse, which is strongly biased by the presence of a wall and does not trace the top of the disk. This single ringed scenario is the bias that \citet{ginski2016} seeks to avoid by only fitting an ellipse to the right side of the disk. However, the full wall is not expected to be illuminated as the presence of the inner ring at 46 au will shadow the wall. The middle panel in Figure \ref{fig:no_innerring} shows a two ringed disk system where the interior ring shadows the outer ring. When we now fit the ellipse to the outer ring in our model, we find that the ellipse traces the true height of the disk. Thus we do not expect to see the effect of a wall in the HD 97048 system.

We fit an ellipse to the second ring of the HD 97048 VLT/SPHERE H-band data. We found the full ellipse of the second ring, we find a minor axis offset of 0$\farcs$1259 $\pm$ 0$\farcs$0013 which, assuming an inclination of 44$^\circ$, results in a scale height of 32 $\pm$ 1.5 au at a projected radius of 172 $\pm$ 4 au. Our ellipse measurements for the second ring can be found in Table \ref{tbl:CUCharings} and shown in Figure \ref{fig:no_innerring}. The presence of a wall can strongly bias the inclination measurement by affecting the ratio of the major to minor axis. Our measured inclination of 44$\pm$3$^\circ$ is consistent with the PAH emission isophot fitting \citep{lagage2006}, and ALMA disk inclination fitting of 41 $\pm$ 3 \citep{walsh2016}. We note that the inclination of 39.9 $\pm$ 1.8$^\circ$ is also consistent with those two alternative measurements. We cannot exclude a wall being present in the HD 97048 data. However, there is no evidence to suggest that the added complexity to the ringed structure is needed. Thus, for this work, we utilize our ellipse fit measurement of the second ring in the HD 97048 system. 

\begin{deluxetable}{lc}
\tablecaption{Gas Disk Parameters}
\tablehead{
\colhead{Parameter} & \colhead{Ring 2}}
\label{tbl:CUCharings}
\startdata
Radius (au) & 172 $\pm$ 4 \\
Inclination ($^\circ$) & 44 $\pm$ 3 \\
Major Axis Offset ($"$) & 0.002 $\pm$ 0.018 \\
Minor Axis Offset ($"$) & 0.1259 $\pm$ 0.0013 \\
Disk Height (au) & 32.3 $\pm$ 1.5 \\
\enddata
\end{deluxetable}


\begin{thebibliography}{}
\expandafter\ifx\csname natexlab\endcsname\relax\def\natexlab#1{#1}\fi
\providecommand{\url}[1]{\href{#1}{#1}}
\providecommand{\dodoi}[1]{doi:~\href{http://doi.org/#1}{\nolinkurl{#1}}}
\providecommand{\doeprint}[1]{\href{http://ascl.net/#1}{\nolinkurl{http://ascl.net/#1}}}
\providecommand{\doarXiv}[1]{\href{https://arxiv.org/abs/#1}{\nolinkurl{https://arxiv.org/abs/#1}}}

\bibitem[{{Andrews} {et~al.}(2018){Andrews}, {Huang}, {P{\'e}rez}, {Isella},
  {Dullemond}, {Kurtovic}, {Guzm{\'a}n}, {Carpenter}, {Wilner}, {Zhang}, {Zhu},
  {Birnstiel}, {Bai}, {Benisty}, {Hughes}, {{\"O}berg}, \&
  {Ricci}}]{andrews2018}
{Andrews}, S.~M., {Huang}, J., {P{\'e}rez}, L.~M., {et~al.} 2018, \apjl, 869,
  L41, \dodoi{10.3847/2041-8213/aaf741}

\bibitem[{{Avenhaus} {et~al.}(2018){Avenhaus}, {Quanz}, {Garufi}, {Perez},
  {Casassus}, {Pinte}, {Bertrang}, {Caceres}, {Benisty}, \&
  {Dominik}}]{avenhaus2018}
{Avenhaus}, H., {Quanz}, S.~P., {Garufi}, A., {et~al.} 2018, \apj, 863, 44,
  \dodoi{10.3847/1538-4357/aab846}

\bibitem[{{Bell} {et~al.}(1997){Bell}, {Cassen}, {Klahr}, \&
  {Henning}}]{bell1997}
{Bell}, K.~R., {Cassen}, P.~M., {Klahr}, H.~H., \& {Henning}, T. 1997, \apj,
  486, 372, \dodoi{10.1086/304514}

\bibitem[{{Bertrang} {et~al.}(2018){Bertrang}, {Avenhaus}, {Casassus},
  {Montesinos}, {Kirchschlager}, {Perez}, {Cieza}, \& {Wolf}}]{bertrang2018}
{Bertrang}, G.~H.~M., {Avenhaus}, H., {Casassus}, S., {et~al.} 2018, \mnras,
  474, 5105, \dodoi{10.1093/mnras/stx3052}

\bibitem[{{Birnstiel} {et~al.}(2010){Birnstiel}, {Dullemond}, \&
  {Brauer}}]{birnstiel2010}
{Birnstiel}, T., {Dullemond}, C.~P., \& {Brauer}, F. 2010, \aap, 513, A79,
  \dodoi{10.1051/0004-6361/200913731}

\bibitem[{{Chiang} \& {Goldreich}(1997)}]{chiang1997}
{Chiang}, E.~I., \& {Goldreich}, P. 1997, \apj, 490, 368,
  \dodoi{10.1086/304869}

\bibitem[{{Cleeves} {et~al.}(2016){Cleeves}, {{\"O}berg}, {Wilner}, {Huang},
  {Loomis}, {Andrews}, \& {Czekala}}]{cleeves2016}
{Cleeves}, L.~I., {{\"O}berg}, K.~I., {Wilner}, D.~J., {et~al.} 2016, \apj,
  832, 110, \dodoi{10.3847/0004-637X/832/2/110}

\bibitem[{{D'Alessio} {et~al.}(1998){D'Alessio}, {Cant{\"o}}, {Calvet}, \&
  {Lizano}}]{dalessio1998}
{D'Alessio}, P., {Cant{\"o}}, J., {Calvet}, N., \& {Lizano}, S. 1998, \apj,
  500, 411, \dodoi{10.1086/305702}

\bibitem[{{Draine} \& {Lee}(1984)}]{draine1984}
{Draine}, B.~T., \& {Lee}, H.~M. 1984, \apj, 285, 89, \dodoi{10.1086/162480}

\bibitem[{{Dullemond} {et~al.}(2001){Dullemond}, {Dominik}, \&
  {Natta}}]{Dullemond2001}
{Dullemond}, C.~P., {Dominik}, C., \& {Natta}, A. 2001, \apj, 560, 957,
  \dodoi{10.1086/323057}

\bibitem[{{Facchini} {et~al.}(2017){Facchini}, {Birnstiel}, {Bruderer}, \& {van
  Dishoeck}}]{facchini2017}
{Facchini}, S., {Birnstiel}, T., {Bruderer}, S., \& {van Dishoeck}, E.~F. 2017,
  \aap, 605, A16, \dodoi{10.1051/0004-6361/201630329}

\bibitem[{{Flaherty} {et~al.}(2017){Flaherty}, {Hughes}, {Rose}, {Simon}, {Qi},
  {Andrews}, {K{\'o}sp{\'a}l}, {Wilner}, {Chiang}, {Armitage}, \&
  {Bai}}]{flaherty2017}
{Flaherty}, K.~M., {Hughes}, A.~M., {Rose}, S.~C., {et~al.} 2017, \apj, 843,
  150, \dodoi{10.3847/1538-4357/aa79f9}

\bibitem[{{Gaia Collaboration} {et~al.}(2016){Gaia Collaboration}, {Prusti},
  {de Bruijne}, {Brown}, {Vallenari}, {Babusiaux}, {Bailer-Jones}, {Bastian},
  {Biermann}, {Evans}, {Eyer}, {Jansen}, {Jordi}, {Klioner}, {Lammers},
  {Lindegren}, {Luri}, {Mignard}, {Milligan}, {Panem}, {Poinsignon},
  {Pourbaix}, {Randich}, {Sarri}, {Sartoretti}, {Siddiqui}, {Soubiran},
  {Valette}, {van Leeuwen}, {Walton}, {Aerts}, {Arenou}, {Cropper}, {Drimmel},
  {H{\o}g}, {Katz}, {Lattanzi}, {O'Mullane}, {Grebel}, {Holland}, {Huc},
  {Passot}, {Bramante}, {Cacciari}, {Casta{\~n}eda}, {Chaoul}, {Cheek}, {De
  Angeli}, {Fabricius}, {Guerra}, {Hern{\'a}ndez}, {Jean-Antoine-Piccolo},
  {Masana}, {Messineo}, {Mowlavi}, {Nienartowicz}, {Ord{\'o}{\~n}ez-Blanco},
  {Panuzzo}, {Portell}, {Richards}, {Riello}, {Seabroke}, {Tanga},
  {Th{\'e}venin}, {Torra}, {Els}, {Gracia-Abril}, {Comoretto},
  {Garcia-Reinaldos}, {Lock}, {Mercier}, {Altmann}, {Andrae}, {Astraatmadja},
  {Bellas-Velidis}, {Benson}, {Berthier}, {Blomme}, {Busso}, {Carry},
  {Cellino}, {Clementini}, {Cowell}, {Creevey}, {Cuypers}, {Davidson}, {De
  Ridder}, {de Torres}, {Delchambre}, {Dell'Oro}, {Ducourant}, {Fr{\'e}mat},
  {Garc{\'\i}a-Torres}, {Gosset}, {Halbwachs}, {Hambly}, {Harrison}, {Hauser},
  {Hestroffer}, {Hodgkin}, {Huckle}, {Hutton}, {Jasniewicz}, {Jordan},
  {Kontizas}, {Korn}, {Lanzafame}, {Manteiga}, {Moitinho}, {Muinonen},
  {Osinde}, {Pancino}, {Pauwels}, {Petit}, {Recio-Blanco}, {Robin}, {Sarro},
  {Siopis}, {Smith}, {Smith}, {Sozzetti}, {Thuillot}, {van Reeven}, {Viala},
  {Abbas}, {Abreu Aramburu}, {Accart}, {Aguado}, {Allan}, {Allasia},
  {Altavilla}, {{\'A}lvarez}, {Alves}, {Anderson}, {Andrei}, {Anglada Varela},
  {Antiche}, {Antoja}, {Ant{\'o}n}, {Arcay}, {Atzei}, {Ayache}, {Bach},
  {Baker}, {Balaguer-N{\'u}{\~n}ez}, {Barache}, {Barata}, {Barbier}, {Barblan},
  {Baroni}, {Barrado y Navascu{\'e}s}, {Barros}, {Barstow}, {Becciani},
  {Bellazzini}, {Bellei}, {Bello Garc{\'\i}a}, {Belokurov}, {Bendjoya},
  {Berihuete}, {Bianchi}, {Bienaym{\'e}}, {Billebaud}, {Blagorodnova},
  {Blanco-Cuaresma}, {Boch}, {Bombrun}, {Borrachero}, {Bouquillon}, {Bourda},
  {Bouy}, {Bragaglia}, {Breddels}, {Brouillet}, {Br{\"u}semeister},
  {Bucciarelli}, {Budnik}, {Burgess}, {Burgon}, {Burlacu}, {Busonero}, {Buzzi},
  {Caffau}, {Cambras}, {Campbell}, {Cancelliere}, {Cantat-Gaudin}, {Carlucci},
  {Carrasco}, {Castellani}, {Charlot}, {Charnas}, {Charvet}, {Chassat},
  {Chiavassa}, {Clotet}, {Cocozza}, {Collins}, {Collins}, {Costigan}, {Crifo},
  {Cross}, {Crosta}, {Crowley}, {Dafonte}, {Damerdji}, {Dapergolas}, {David},
  {David}, {De Cat}, {de Felice}, {de Laverny}, {De Luise}, {De March}, {de
  Martino}, {de Souza}, {Debosscher}, {del Pozo}, {Delbo}, {Delgado},
  {Delgado}, {di Marco}, {Di Matteo}, {Diakite}, {Distefano}, {Dolding}, {Dos
  Anjos}, {Drazinos}, {Dur{\'a}n}, {Dzigan}, {Ecale}, {Edvardsson}, {Enke},
  {Erdmann}, {Escolar}, {Espina}, {Evans}, {Eynard Bontemps}, {Fabre},
  {Fabrizio}, {Faigler}, {Falc{\~a}o}, {Farr{\`a}s Casas}, {Faye}, {Federici},
  {Fedorets}, {Fern{\'a}ndez-Hern{\'a}ndez}, {Fernique}, {Fienga}, {Figueras},
  {Filippi}, {Findeisen}, {Fonti}, {Fouesneau}, {Fraile}, {Fraser}, {Fuchs},
  {Furnell}, {Gai}, {Galleti}, {Galluccio}, {Garabato}, {Garc{\'\i}a-Sedano},
  {Gar{\'e}}, {Garofalo}, {Garralda}, {Gavras}, {Gerssen}, {Geyer}, {Gilmore},
  {Girona}, {Giuffrida}, {Gomes}, {Gonz{\'a}lez-Marcos},
  {Gonz{\'a}lez-N{\'u}{\~n}ez}, {Gonz{\'a}lez-Vidal}, {Granvik}, {Guerrier},
  {Guillout}, {Guiraud}, {G{\'u}rpide}, {Guti{\'e}rrez-S{\'a}nchez}, {Guy},
  {Haigron}, {Hatzidimitriou}, {Haywood}, {Heiter}, {Helmi}, {Hobbs},
  {Hofmann}, {Holl}, {Holland }, {Hunt}, {Hypki}, {Icardi}, {Irwin}, {Jevardat
  de Fombelle}, {Jofr{\'e}}, {Jonker}, {Jorissen}, {Julbe}, {Karampelas},
  {Kochoska}, {Kohley}, {Kolenberg}, {Kontizas}, {Koposov}, {Kordopatis},
  {Koubsky}, {Kowalczyk}, {Krone-Martins}, {Kudryashova}, {Kull}, {Bachchan},
  {Lacoste-Seris}, {Lanza}, {Lavigne}, {Le Poncin-Lafitte}, {Lebreton},
  {Lebzelter}, {Leccia}, {Leclerc}, {Lecoeur-Taibi}, {Lemaitre}, {Lenhardt},
  {Leroux}, {Liao}, {Licata}, {Lindstr{\o}m}, {Lister}, {Livanou}, {Lobel},
  {L{\"o}ffler}, {L{\'o}pez}, {Lopez-Lozano}, {Lorenz}, {Loureiro},
  {MacDonald}, {Magalh{\~a}es Fernandes}, {Managau}, {Mann}, {Mantelet},
  {Marchal}, {Marchant}, {Marconi}, {Marie}, {Marinoni}, {Marrese},
  {Marschalk{\'o}}, {Marshall}, {Mart{\'\i}n-Fleitas}, {Martino}, {Mary},
  {Matijevi{\v{c}}}, {Mazeh}, {McMillan}, {Messina}, {Mestre}, {Michalik},
  {Millar}, {Miranda}, {Molina}, {Molinaro}, {Molinaro}, {Moln{\'a}r},
  {Moniez}, {Montegriffo}, {Monteiro}, {Mor}, {Mora}, {Morbidelli}, {Morel},
  {Morgenthaler}, {Morley}, {Morris}, {Mulone}, {Muraveva}, {Musella},
  {Narbonne}, {Nelemans}, {Nicastro}, {Noval}, {Ord{\'e}novic},
  {Ordieres-Mer{\'e}}, {Osborne}, {Pagani}, {Pagano}, {Pailler}, {Palacin},
  {Palaversa}, {Parsons}, {Paulsen}, {Pecoraro}, {Pedrosa}, {Pentik{\"a}inen},
  {Pereira}, {Pichon}, {Piersimoni}, {Pineau}, {Plachy}, {Plum}, {Poujoulet},
  {Pr{\v{s}}a}, {Pulone}, {Ragaini}, {Rago}, {Rambaux}, {Ramos-Lerate},
  {Ranalli}, {Rauw}, {Read}, {Regibo}, {Renk}, {Reyl{\'e}}, {Ribeiro},
  {Rimoldini}, {Ripepi}, {Riva}, {Rixon}, {Roelens}, {Romero-G{\'o}mez},
  {Rowell}, {Royer}, {Rudolph}, {Ruiz-Dern}, {Sadowski}, {Sagrist{\`a}
  Sell{\'e}s}, {Sahlmann}, {Salgado}, {Salguero}, {Sarasso}, {Savietto},
  {Schnorhk}, {Schultheis}, {Sciacca}, {Segol}, {Segovia}, {Segransan},
  {Serpell}, {Shih}, {Smareglia}, {Smart}, {Smith}, {Solano}, {Solitro},
  {Sordo}, {Soria Nieto}, {Souchay}, {Spagna}, {Spoto}, {Stampa}, {Steele},
  {Steidelm{\"u}ller}, {Stephenson}, {Stoev}, {Suess}, {S{\"u}veges}, {Surdej},
  {Szabados}, {Szegedi-Elek}, {Tapiador}, {Taris}, {Tauran}, {Taylor},
  {Teixeira}, {Terrett}, {Tingley}, {Trager}, {Turon}, {Ulla}, {Utrilla},
  {Valentini}, {van Elteren}, {Van Hemelryck}, {van Leeuwen}, {Varadi},
  {Vecchiato}, {Veljanoski}, {Via}, {Vicente}, {Vogt}, {Voss}, {Votruba},
  {Voutsinas}, {Walmsley}, {Weiler}, {Weingrill}, {Werner}, {Wevers},
  {Whitehead}, {Wyrzykowski}, {Yoldas}, {{\v{Z}}erjal}, {Zucker}, {Zurbach},
  {Zwitter}, {Alecu}, {Allen}, {Allende Prieto}, {Amorim},
  {Anglada-Escud{\'e}}, {Arsenijevic}, {Azaz}, {Balm}, {Beck}, {Bernstein},
  {Bigot}, {Bijaoui}, {Blasco}, {Bonfigli}, {Bono}, {Boudreault}, {Bressan},
  {Brown}, {Brunet}, {Bunclark}, {Buonanno}, {Butkevich}, {Carret}, {Carrion},
  {Chemin}, {Ch{\'e}reau}, {Corcione}, {Darmigny}, {de Boer}, {de Teodoro}, {de
  Zeeuw}, {Delle Luche}, {Domingues}, {Dubath}, {Fodor}, {Fr{\'e}zouls},
  {Fries}, {Fustes}, {Fyfe}, {Gallardo}, {Gallegos}, {Gardiol}, {Gebran},
  {Gomboc}, {G{\'o}mez}, {Grux}, {Gueguen}, {Heyrovsky}, {Hoar}, {Iannicola},
  {Isasi Parache}, {Janotto}, {Joliet}, {Jonckheere}, {Keil}, {Kim},
  {Klagyivik}, {Klar}, {Knude}, {Kochukhov}, {Kolka}, {Kos}, {Kutka}, {Lainey},
  {LeBouquin}, {Liu}, {Loreggia}, {Makarov}, {Marseille}, {Martayan},
  {Martinez-Rubi}, {Massart}, {Meynadier}, {Mignot}, {Munari}, {Nguyen},
  {Nordlander}, {Ocvirk}, {O'Flaherty}, {Olias Sanz}, {Ortiz}, {Osorio},
  {Oszkiewicz}, {Ouzounis}, {Palmer}, {Park}, {Pasquato}, {Peltzer}, {Peralta},
  {P{\'e}turaud}, {Pieniluoma}, {Pigozzi}, {Poels}, {Prat}, {Prod'homme},
  {Raison}, {Rebordao}, {Risquez}, {Rocca-Volmerange}, {Rosen}, {Ruiz-Fuertes},
  {Russo}, {Sembay}, {Serraller Vizcaino}, {Short}, {Siebert}, {Silva},
  {Sinachopoulos}, {Slezak}, {Soffel}, {Sosnowska}, {Strai{\v{z}}ys}, {ter
  Linden}, {Terrell}, {Theil}, {Tiede}, {Troisi}, {Tsalmantza}, {Tur},
  {Vaccari}, {Vachier}, {Valles}, {Van Hamme}, {Veltz}, {Virtanen}, {Wallut},
  {Wichmann}, {Wilkinson}, {Ziaeepour}, \& {Zschocke}}]{gaia2016}
{Gaia Collaboration}, {Prusti}, T., {de Bruijne}, J.~H.~J., {et~al.} 2016,
  \aap, 595, A1, \dodoi{10.1051/0004-6361/201629272}

\bibitem[{{Gaia Collaboration} {et~al.}(2020){Gaia Collaboration}, {Smart},
  {Sarro}, {Rybizki}, {Reyl{\'e}}, {Robin}, {Hambly}, {Abbas}, {Barstow}, {de
  Bruijne}, {Bucciarelli}, {Carrasco}, {Cooper}, {Hodgkin}, {Masana},
  {Michalik}, {Sahlmann}, {Sozzetti}, {Brown}, {Vallenari}, {Prusti},
  {Babusiaux}, {Biermann}, {Creevey}, {Evans}, {Eyer}, {Hutton}, {Jansen},
  {Jordi}, {Klioner}, {Lammers}, {Lindegren}, {Luri}, {Mignard}, {Panem},
  {Pourbaix}, {Randich}, {Sartoretti}, {Soubiran}, {Walton}, {Arenou},
  {Bailer-Jones}, {Bastian}, {Cropper}, {Drimmel}, {Katz}, {Lattanzi}, {van
  Leeuwen}, {Bakker}, {Casta{\~n}eda}, {De Angeli}, {Ducourant}, {Fabricius},
  {Fouesneau}, {Fr{\'e}mat}, {Guerra}, {Guerrier}, {Guiraud}, {Jean-Antoine
  Piccolo}, {Messineo}, {Mowlavi}, {Nicolas}, {Nienartowicz}, {Pailler},
  {Panuzzo}, {Riclet}, {Roux}, {Seabroke}, {Sordo}, {Tanga}, {Th{\'e}venin},
  {Gracia-Abril}, {Portell}, {Teyssier}, {Altmann}, {Andrae}, {Bellas-Velidis},
  {Benson}, {Berthier}, {Blomme}, {Brugaletta}, {Burgess}, {Busso}, {Carry},
  {Cellino}, {Cheek}, {Clementini}, {Damerdji}, {Davidson}, {Delchambre},
  {Dell'Oro}, {Fern{\'a}ndez-Hern{\'a}ndez}, {Galluccio}, {Garc{\'\i}a-Lario},
  {Garcia-Reinaldos}, {Gonz{\'a}lez-N{\'u}{\~n}ez}, {Gosset}, {Haigron},
  {Halbwachs}, {Harrison}, {Hatzidimitriou}, {Heiter}, {Hern{\'a}ndez},
  {Hestroffer}, {Holl}, {Jan{\ss}en}, {Jevardat de Fombelle}, {Jordan},
  {Krone-Martins}, {Lanzafame}, {L{\"o}ffler}, {Lorca}, {Manteiga}, {Marchal},
  {Marrese}, {Moitinho}, {Mora}, {Muinonen}, {Osborne}, {Pancino}, {Pauwels},
  {Recio-Blanco}, {Richards}, {Riello}, {Rimoldini}, {Roegiers}, {Siopis},
  {Smith}, {Ulla}, {Utrilla}, {van Leeuwen}, {van Reeven}, {Abreu Aramburu},
  {Accart}, {Aerts}, {Aguado}, {Ajaj}, {Altavilla}, {{\'A}lvarez}, {{\'A}lvarez
  Cid-Fuentes}, {Alves}, {Anderson}, {Anglada Varela}, {Antoja}, {Audard},
  {Baines}, {Baker}, {Balaguer-N{\'u}{\~n}ez}, {Balbinot}, {Balog}, {Barache},
  {Barbato}, {Barros}, {Bartolom{\'e}}, {Bassilana}, {Bauchet},
  {Baudesson-Stella}, {Becciani}, {Bellazzini}, {Bernet}, {Bertone}, {Bianchi},
  {Blanco-Cuaresma}, {Boch}, {Bombrun}, {Bossini}, {Bouquillon}, {Bragaglia},
  {Bramante}, {Breedt}, {Bressan}, {Brouillet}, {Burlacu}, {Busonero},
  {Butkevich}, {Buzzi}, {Caffau}, {Cancelliere}, {C{\'a}novas},
  {Cantat-Gaudin}, {Carballo}, {Carlucci}, {Carnerero}, {Casamiquela},
  {Castellani}, {Castro-Ginard}, {Castro Sampol}, {Chaoul}, {Charlot},
  {Chemin}, {Chiavassa}, {Cioni}, {Comoretto}, {Cornez}, {Cowell}, {Crifo},
  {Crosta}, {Crowley}, {Dafonte}, {Dapergolas}, {David}, {David}, {de Laverny},
  {De Luise}, {De March}, {De Ridder}, {de Souza}, {de Teodoro}, {de Torres},
  {del Peloso}, {del Pozo}, {Delgado}, {Delgado}, {Delisle}, {Di Matteo},
  {Diakite}, {Diener}, {Distefano}, {Dolding}, {Eappachen}, {Edvardsson},
  {Enke}, {Esquej}, {Fabre}, {Fabrizio}, {Faigler}, {Fedorets}, {Fernique},
  {Fienga}, {Figueras}, {Fouron}, {Fragkoudi}, {Fraile}, {Franke}, {Gai},
  {Garabato}, {Garcia-Gutierrez}, {Garc{\'\i}a-Torres}, {Garofalo}, {Gavras},
  {Gerlach}, {Geyer}, {Giacobbe}, {Gilmore}, {Girona}, {Giuffrida}, {Gomel},
  {Gomez}, {Gonzalez-Santamaria}, {Gonz{\'a}lez-Vidal}, {Granvik},
  {Guti{\'e}rrez-S{\'a}nchez}, {Guy}, {Hauser}, {Haywood}, {Helmi}, {Hidalgo},
  {Hilger}, {H{\l}adczuk}, {Hobbs}, {Holland}, {Huckle}, {Jasniewicz},
  {Jonker}, {Juaristi Campillo}, {Julbe}, {Karbevska}, {Kervella}, {Khanna},
  {Kochoska}, {Kontizas}, {Kordopatis}, {Korn}, {Kostrzewa-Rutkowska},
  {Kruszy{\'n}ska}, {Lambert}, {Lanza}, {Lasne}, {Le Campion}, {Le Fustec},
  {Lebreton}, {Lebzelter}, {Leccia}, {Leclerc}, {Lecoeur-Taibi}, {Liao},
  {Licata}, {Lindstr{\o}m}, {Lister}, {Livanou}, {Lobel}, {Madrero Pardo},
  {Managau}, {Mann}, {Marchant}, {Marconi}, {Marcos Santos}, {Marinoni},
  {Marocco}, {Marshall}, {Polo}, {Mart{\'\i}n-Fleitas}, {Masip}, {Massari},
  {Mastrobuono-Battisti}, {Mazeh}, {McMillan}, {Messina}, {Millar}, {Mints},
  {Molina}, {Molinaro}, {Moln{\'a}r}, {Montegriffo}, {Mor}, {Morbidelli},
  {Morel}, {Morris}, {Mulone}, {Munoz}, {Muraveva}, {Murphy}, {Musella},
  {Noval}, {Ord{\'e}novic}, {Orr{\`u}}, {Osinde}, {Pagani}, {Pagano},
  {Palaversa}, {Palicio}, {Panahi}, {Pawlak}, {Pe{\~n}alosa Esteller},
  {Penttil{\"a}}, {Piersimoni}, {Pineau}, {Plachy}, {Plum}, {Poggio},
  {Poretti}, {Poujoulet}, {Pr{\v{s}}a}, {Pulone}, {Racero}, {Ragaini},
  {Rainer}, {Raiteri}, {Rambaux}, {Ramos}, {Ramos-Lerate}, {Re Fiorentin},
  {Regibo}, {Ripepi}, {Riva}, {Rixon}, {Robichon}, {Robin}, {Roelens},
  {Rohrbasser}, {Romero-G{\'o}mez}, {Rowell}, {Royer}, {Rybicki}, {Sadowski},
  {Sagrist{\`a} Sell{\'e}s}, {Salgado}, {Salguero}, {Samaras}, {Sanchez
  Gimenez}, {Sanna}, {Santove{\~n}a}, {Sarasso}, {Schultheis}, {Sciacca},
  {Segol}, {Segovia}, {S{\'e}gransan}, {Semeux}, {Shahaf}, {Siddiqui},
  {Siebert}, {Siltala}, {Slezak}, {Solano}, {Solitro}, {Souami}, {Souchay},
  {Spagna}, {Spoto}, {Steele}, {Steidelm{\"u}ller}, {Stephenson},
  {S{\"u}veges}, {Szabados}, {Szegedi-Elek}, {Taris}, {Tauran}, {Taylor},
  {Teixeira}, {Thuillot}, {Tonello}, {Torra}, {Torra}, {Turon}, {Unger},
  {Vaillant}, {van Dillen}, {Vanel}, {Vecchiato}, {Viala}, {Vicente},
  {Voutsinas}, {Weiler}, {Wevers}, {Wyrzykowski}, {Yoldas}, {Yvard}, {Zhao},
  {Zorec}, {Zucker}, {Zurbach}, \& {Zwitter}}]{gaia2020}
{Gaia Collaboration}, {Smart}, R.~L., {Sarro}, L.~M., {et~al.} 2020, arXiv
  e-prints, arXiv:2012.02061.
\newblock \doarXiv{2012.02061}

\bibitem[{{Ginski} {et~al.}(2016){Ginski}, {Stolker}, {Pinilla}, {Dominik},
  {Boccaletti}, {de Boer}, {Benisty}, {Biller}, {Feldt}, {Garufi}, {Keller},
  {Kenworthy}, {Maire}, {M{\'e}nard}, {Mesa}, {Milli}, {Min}, {Pinte}, {Quanz},
  {van Boekel}, {Bonnefoy}, {Chauvin}, {Desidera}, {Gratton}, {Girard},
  {Keppler}, {Kopytova}, {Lagrange}, {Langlois}, {Rouan}, \&
  {Vigan}}]{ginski2016}
{Ginski}, C., {Stolker}, T., {Pinilla}, P., {et~al.} 2016, \aap, 595, A112,
  \dodoi{10.1051/0004-6361/201629265}

\bibitem[{{Harries}(2000)}]{harries2000}
{Harries}, T.~J. 2000, \mnras, 315, 722,
  \dodoi{10.1046/j.1365-8711.2000.03505.x}

\bibitem[{{Harries}(2011)}]{harries2011}
---. 2011, \mnras, 416, 1500, \dodoi{10.1111/j.1365-2966.2011.19147.x}

\bibitem[{{Harries} {et~al.}(2004){Harries}, {Monnier}, {Symington}, \&
  {Kurosawa}}]{harries2004}
{Harries}, T.~J., {Monnier}, J.~D., {Symington}, N.~H., \& {Kurosawa}, R. 2004,
  \mnras, 350, 565, \dodoi{10.1111/j.1365-2966.2004.07668.x}

\bibitem[{{Hughes} {et~al.}(2011){Hughes}, {Wilner}, {Andrews}, {Qi}, \&
  {Hogerheijde}}]{hughes2011}
{Hughes}, A.~M., {Wilner}, D.~J., {Andrews}, S.~M., {Qi}, C., \& {Hogerheijde},
  M.~R. 2011, \apj, 727, 85, \dodoi{10.1088/0004-637X/727/2/85}

\bibitem[{{Hughes} {et~al.}(1994){Hughes}, {Hartigan}, {Krautter}, \&
  {Kelemen}}]{hughes1994}
{Hughes}, J., {Hartigan}, P., {Krautter}, J., \& {Kelemen}, J. 1994, \aj, 108,
  1071, \dodoi{10.1086/117135}

\bibitem[{{Isella} {et~al.}(2016){Isella}, {Guidi}, {Testi}, {Liu}, {Li}, {Li},
  {Weaver}, {Boehler}, {Carperter}, {De Gregorio-Monsalvo}, {Manara}, {Natta},
  {P{\'e}rez}, {Ricci}, {Sargent}, {Tazzari}, \& {Turner}}]{isella2016}
{Isella}, A., {Guidi}, G., {Testi}, L., {et~al.} 2016, \prl, 117, 251101,
  \dodoi{10.1103/PhysRevLett.117.251101}

\bibitem[{{Kenyon} \& {Hartmann}(1987)}]{k1987}
{Kenyon}, S.~J., \& {Hartmann}, L. 1987, \apj, 323, 714, \dodoi{10.1086/165866}

\bibitem[{{Kraus} {et~al.}(2020){Kraus}, {Kreplin}, {Young}, {Bate}, {Monnier},
  {Harries}, {Avenhaus}, {Kluska}, {Laws}, {Rich}, {Willson}, {Aarnio},
  {Adams}, {Andrews}, {Anugu}, {Bae}, {ten Brummelaar}, {Calvet}, {Cur{\'e}},
  {Davies}, {Ennis}, {Espaillat}, {Gardner}, {Hartmann}, {Hinkley}, {Labdon},
  {Lanthermann}, {LeBouquin}, {Schaefer}, {Setterholm}, {Wilner}, \&
  {Zhu}}]{kraus2020}
{Kraus}, S., {Kreplin}, A., {Young}, A.~K., {et~al.} 2020, Science, 369, 1233,
  \dodoi{10.1126/science.aba4633}

\bibitem[{{Lagage} {et~al.}(2006){Lagage}, {Doucet}, {Pantin}, {Habart},
  {Duch{\^e}ne}, {M{\'e}nard}, {Pinte}, {Charnoz}, \& {Pel}}]{lagage2006}
{Lagage}, P.-O., {Doucet}, C., {Pantin}, E., {et~al.} 2006, Science, 314, 621,
  \dodoi{10.1126/science.1131436}

\bibitem[{{Lucy}(1999)}]{lucy1999}
{Lucy}, L.~B. 1999, \aap, 344, 282

\bibitem[{{Malbet} \& {Bertout}(1991)}]{fabien1991}
{Malbet}, F., \& {Bertout}, C. 1991, \apj, 383, 814, \dodoi{10.1086/170839}

\bibitem[{{Manoj} {et~al.}(2006){Manoj}, {Bhatt}, {Maheswar}, \&
  {Muneer}}]{Manoj2006}
{Manoj}, P., {Bhatt}, H.~C., {Maheswar}, G., \& {Muneer}, S. 2006, \apj, 653,
  657, \dodoi{10.1086/508764}

\bibitem[{{Meyer} \& {Meyer-Hofmeister}(1982)}]{meyer1982}
{Meyer}, F., \& {Meyer-Hofmeister}, E. 1982, \aap, 106, 34

\bibitem[{{Monnier} {et~al.}(2017){Monnier}, {Harries}, {Aarnio}, {Adams},
  {Andrews}, {Calvet}, {Espaillat}, {Hartmann}, {Hinkley}, {Kraus}, {McClure},
  {Oppenheimer}, {Perrin}, \& {Wilner}}]{monnier2017}
{Monnier}, J.~D., {Harries}, T.~J., {Aarnio}, A., {et~al.} 2017, \apj, 838, 20,
  \dodoi{10.3847/1538-4357/aa6248}

\bibitem[{{Montesinos} {et~al.}(2021){Montesinos}, {Cuello}, {Olofsson},
  {Cuadra}, {Bayo}, {Bertrang}, \& {Perrot}}]{Montesinos2021}
{Montesinos}, M., {Cuello}, N., {Olofsson}, J., {et~al.} 2021, \apj, 910, 31,
  \dodoi{10.3847/1538-4357/abe3fc}

\bibitem[{{M{\"u}ller} {et~al.}(2005){M{\"u}ller}, {Schl{\"o}der}, {Stutzki},
  \& \t{Winnewisser}}]{muller2005}
{M{\"u}ller}, H.~S.~P., {Schl{\"o}der}, F., {Stutzki}, J., \& \t{Winnewisser},
  G. 2005, Journal of Molecular Structure, 742, 215,
  \dodoi{10.1016/j.molstruc.2005.01.027}

\bibitem[{{Natta} {et~al.}(2001){Natta}, {Prusti}, {Neri}, {Wooden}, {Grinin},
  \& {Mannings}}]{natta2001}
{Natta}, A., {Prusti}, T., {Neri}, R., {et~al.} 2001, \aap, 371, 186,
  \dodoi{10.1051/0004-6361:20010334}

\bibitem[{{Pani{\'c}} {et~al.}(2009){Pani{\'c}}, {Hogerheijde}, {Wilner}, \&
  {Qi}}]{panic2009}
{Pani{\'c}}, O., {Hogerheijde}, M.~R., {Wilner}, D., \& {Qi}, C. 2009, \aap,
  501, 269, \dodoi{10.1051/0004-6361/200911883}

\bibitem[{{Pinte} {et~al.}(2018){Pinte}, {M{\'e}nard}, {Duch{\^e}ne}, {Hill},
  {Dent}, {Woitke}, {Maret}, {van der Plas}, {Hales}, {Kamp}, {Thi}, {de
  Gregorio-Monsalvo}, {Rab}, {Quanz}, {Avenhaus}, {Carmona}, \&
  {Casassus}}]{pinte2018}
{Pinte}, C., {M{\'e}nard}, F., {Duch{\^e}ne}, G., {et~al.} 2018, \aap, 609,
  A47, \dodoi{10.1051/0004-6361/201731377}

\bibitem[{{Podio} {et~al.}(2020){Podio}, {Garufi}, {Codella}, {Fedele},
  {Bianchi}, {Bacciotti}, {Ceccarelli}, {Favre}, {Mercimek}, {Rygl}, \&
  {Testi}}]{podio2020}
{Podio}, L., {Garufi}, A., {Codella}, C., {et~al.} 2020, \aap, 642, L7,
  \dodoi{10.1051/0004-6361/202038952}

\bibitem[{{Pohl} {et~al.}(2017){Pohl}, {Benisty}, {Pinilla}, {Ginski}, {de
  Boer}, {Avenhaus}, {Henning}, {Zurlo}, {Boccaletti}, {Augereau}, {Birnstiel},
  {Dominik}, {Facchini}, {Fedele}, {Janson}, {Keppler}, {Kral}, {Langlois},
  {Ligi}, {Maire}, {M{\'e}nard}, {Meyer}, {Pinte}, {Quanz}, {Sauvage},
  {Sezestre}, {Stolker}, {Szul{\'a}gyi}, {van Boekel}, {van der Plas},
  {Villenave}, {Baruffolo}, {Baudoz}, {Le Mignant}, {Maurel}, {Ramos}, \&
  {Weber}}]{pohl2017}
{Pohl}, A., {Benisty}, M., {Pinilla}, P., {et~al.} 2017, \apj, 850, 52,
  \dodoi{10.3847/1538-4357/aa94c2}

\bibitem[{{Rich} {et~al.}(2020){Rich}, {Wisniewski}, {Sitko}, \&
  {Grady}}]{rich2020}
{Rich}, E.~A., {Wisniewski}, J.~P., {Sitko}, M., \& {Grady}, C. 2020, Submitted

\bibitem[{{Rich} {et~al.}(2019){Rich}, {Wisniewski}, {Currie}, {Fukagawa},
  {Grady}, {Sitko}, {Pikhartova}, {Hashimoto}, {Abe}, {Brandner}, {Brandt},
  {Carson}, {Chilcote}, {Dong}, {Feldt}, {Goto}, {Groff}, {Guyon}, {Hayano},
  {Hayashi}, {Hayashi}, {Henning}, {Hodapp}, {Ishii}, {Iye}, {Janson},
  {Jovanovic}, {Kandori}, {Kasdin}, {Knapp}, {Kudo}, {Kusakabe}, {Kuzuhara},
  {Kwon}, {Lozi}, {Martinache}, {Matsuo}, {Mayama}, {McElwain}, {Miyama},
  {Morino}, {Moro-Martin}, {Nakagawa}, {Nishimura}, {Pyo}, {Serabyn}, {Suto},
  {Russel}, {Suzuki}, {Takami}, {Takato}, {Terada}, {Thalmann}, {Turner},
  {Uyama}, {Wagner}, {Watanabe}, {Yamada}, {Takami}, {Usuda}, \&
  {Tamura}}]{rich2019}
{Rich}, E.~A., {Wisniewski}, J.~P., {Currie}, T., {et~al.} 2019, \apj, 875, 38,
  \dodoi{10.3847/1538-4357/ab0f3b}

\bibitem[{{Rundle} {et~al.}(2010){Rundle}, {Harries}, {Acreman}, \&
  {Bate}}]{rundle2010}
{Rundle}, D., {Harries}, T.~J., {Acreman}, D.~M., \& {Bate}, M.~R.~t. 2010,
  \\mnras, 407, 986, \dodoi{10.1111/j.1365-2966.2010.16982.x}

\bibitem[{{Sch{\"o}ier} {et~al.}(2005){Sch{\"o}ier}, {van der Tak}, {van
  Dishoeck}, \& \t{Black}}]{schoier2005}
{Sch{\"o}ier}, F.~L., {van der Tak}, F.~F.~S., {van Dishoeck}, E.~F., \&
  \t{Black}, J.~H. 2005, \aap, 432, 369, \dodoi{10.1051/0004-6361:20041729}

\bibitem[{{Teague}(2019)}]{teague2019}
{Teague}, R. 2019, The Journal of Open Source Software, 4, 1632,
  \dodoi{10.21105/joss.01632}

\bibitem[{{Teague} {et~al.}(2018){Teague}, {Bae}, {Bergin}, {Birnstiel}, \&
  {Foreman-Mackey}}]{teague2018}
{Teague}, R., {Bae}, J., {Bergin}, E.~A., {Birnstiel}, T., \& {Foreman-Mackey},
  D. 2018, \apjl, 860, L12, \dodoi{10.3847/2041-8213/aac6d7}

\bibitem[{{van den Ancker} {et~al.}(1998){van den Ancker}, {de Winter}, \&
  {Tjin A Djie}}]{ancker1998}
{van den Ancker}, M.~E., {de Winter}, D., \& {Tjin A Djie}, H.~R.~E. 1998,
  \aap, 330, 145

\bibitem[{{Villenave} {et~al.}(2020){Villenave}, {M{\'e}nard}, {Dent},
  {Duch{\^e}ne}, {Stapelfeldt}, {Benisty}, {Boehler}, {van der Plas}, {Pinte},
  {Telkamp}, {Wolff}, {Flores}, {Lesur}, {Louvet}, {Riols}, {Dougados},
  {Williams}, \& {Padgett}}]{villenave2020}
{Villenave}, M., {M{\'e}nard}, F., {Dent}, W.~R.~F., {et~al.} 2020, \aap, 642,
  A164, \dodoi{10.1051/0004-6361/202038087}

\bibitem[{{Vioque} {et~al.}(2018){Vioque}, {Oudmaijer}, {Baines},
  {Mendigut{\'\i}a}, \& {P{\'e}rez-Mart{\'\i}nez}}]{vioque2018}
{Vioque}, M., {Oudmaijer}, R.~D., {Baines}, D., {Mendigut{\'\i}a}, I., \&
  {P{\'e}rez-Mart{\'\i}nez}, R. 2018, \aap, 620, A128,
  \dodoi{10.1051/0004-6361/201832870}

\bibitem[{{Walsh} {et~al.}(2016){Walsh}, {Juh{\'a}sz}, {Meeus}, {Dent}, {Maud},
  {Aikawa}, {Millar}, \& {Nomura}}]{walsh2016}
{Walsh}, C., {Juh{\'a}sz}, A., {Meeus}, G., {et~al.} 2016, \apj, 831, 200,
  \dodoi{10.3847/0004-637X/831/2/200}

\bibitem[{{Wichittanakom} {et~al.}(2020){Wichittanakom}, {Oudmaijer},
  {Fairlamb}, {Mendigut{\'\i}a}, {Vioque}, \& {Ababakr}}]{wichittanakom2020}
{Wichittanakom}, C., {Oudmaijer}, R.~D., {Fairlamb}, J.~R., {et~al.} 2020,
  \mnras, 493, 234, \dodoi{10.1093/mnras/staa169}

\bibitem[{{Williams} \& {Best}(2014)}]{williams2014}
{Williams}, J.~P., \& {Best}, W. M.~J. 2014, \apj, 788, 59,
  \dodoi{10.1088/0004-637X/788/1/59}

\bibitem[{{Wisniewski} {et~al.}(2008){Wisniewski}, {Clampin}, {Grady},
  {Ardila}, {Ford}, {Golimowski}, {Illingworth}, \& {Krist}}]{wisniewski2008}
{Wisniewski}, J.~P., {Clampin}, M., {Grady}, C.~A., {et~al.} 2008, \apj, 682,
  548, \dodoi{10.1086/589629}

\bibitem[{{Yang} {et~al.}(2010){Yang}, {Stancil}, {Balakrishnan}, \&
  {Forrey}}]{yang2010}
{Yang}, B., {Stancil}, P.~C., {Balakrishnan}, N., \& {Forrey}, R.~C.~t. 2010,
  \apj, 718, 1062, \dodoi{10.1088/0004-637X/718/2/1062}

\end{thebibliography}

\end{document}